\documentclass[iop]{emulateapj}
\usepackage{amssymb,multirow,color}

\usepackage{bm}
\usepackage{epsfig}
\usepackage{graphicx}
\usepackage{color}
\usepackage{subfigure}

\newcommand{\be}{\begin{equation}}
\newcommand{\ee}{\end{equation}}
\newcommand{\beq}{\begin{equation}}
\newcommand{\eeq}{\end{equation}}

\newcommand\bea{\begin{eqnarray}}
\newcommand\eea{\end{eqnarray}}

\begin{document}
\title{Magnetohydrodynamic turbulence mediated by reconnection}
\author{Stanislav Boldyrev$^{1,2}$, Nuno\ F.\ Loureiro$^{3}$}
\affil{${~}^1$Department of Physics, University of Wisconsin, Madison, WI 53706, USA\\
${~}^2$Space Science Institute, Boulder, Colorado 80301, USA\\
${~}^3$Plasma Science and Fusion Center, Massachusetts Institute of Technology, Cambridge MA 02139, USA}

\date{\em Submitted to Astrophys. J. May 09, 2017}

\begin{abstract}
Magnetic field fluctuations in MHD turbulence can be viewed as current sheets that are progressively more anisotropic at smaller scales. As suggested by \citet{loureiro2017} and \citet{mallet2017}, below a certain critical thickness $\lambda_c$ such current sheets become tearing-unstable. We propose that the tearing instability changes the effective alignment of the magnetic field lines in such a way as to balance the eddy turnover rate at all scales smaller than $\lambda_c$. As a result, turbulent fluctuations become progressively less anisotropic at smaller scales, with the alignment angle increasing as $\theta \sim (\lambda/\lambda_*)^{-4/5+\beta}$, where $\lambda_*\sim L_0 S_0^{-3/4}$ is the resistive dissipation scale. Here $L_0$ is the outer scale of the turbulence, $S_0$ is the corresponding Lundquist number, and {$0\leq \beta <4/5$} is a parameter. The resulting Fourier energy spectrum is $E(k_\perp)\propto k_\perp^{-11/5+2\beta/3}$, where $k_\perp$ is the wavenumber normal to the local mean magnetic field, and the critical scale is $\lambda_c\sim S_L^{-(4-5\beta)/(7-{20\beta/3})}$. The simplest model corresponds to $\beta=0$, in which case the predicted scaling formally agrees with one of the solutions obtained in \cite[][]{mallet2017} from a discrete hierarchical model of abruptly collapsing current sheets, an approach different and complementary to ours. We also show that the reconnection-mediated interval is non-universal with respect to the dissipation mechanism. Hyper-resistivity of the form ${\tilde \eta}k^{2+2s}$ leads (in the simplest case of $\beta=0$) to the different transition scale $\lambda_c\sim L_0{\tilde S}_0^{-4/(7+9s)}$ and the energy spectrum $E(k_\perp)\propto k_\perp^{-(11+9s)/(5+3s)}$, where ${\tilde S}_0$ is the corresponding hyper-resistive Lundquist number. 
\end{abstract}
\maketitle


\section{Introduction}
Magnetohydrodynamics (MHD) turbulence plays an important role in a variety of astrophysical phenomena, including convective flows and dynamo action in stars, angular momentum transport in accretion discs, heating of stellar coronae and winds, generation of structures in the interstellar medium, heat conduction in galaxy clusters, etc \cite[e.g.,][]{biskamp2003,elmegreen2004,tobias2013}.   Recently it became clear that the understanding of MHD turbulence is incomplete without understanding the role that magnetic reconnection plays in a turbulent cascade. Indeed, MHD turbulent structures at small scales look like current sheets \cite[e.g.,][]{matthaeus_turbulent_1986,servidio_magnetic_2009,servidio_magnetic_2011,biskamp2003,wan2013,zhdankin_etal2013,zhdankin_etal2014}. On the other hand, as current sheets reconnect due to tearing instability, they generate small-scale turbulence inside them that is qualitatively different from the standard Alfv\'enic turbulence \cite[e.g.,][]{loureiro_instability_2007,lapenta_self_2008,samtaney_formation_2009,bhattacharjee_fast_2009,uzdensky_fast_2010,huang_scaling_2010,barta_spontaneous_2011,loureiro_magnetic_2012,loureiro2013,huang_turbulent_2016,loureiro_magnetic_2016}. In \cite[][]{loureiro2017,mallet2017} it was suggested that the energy cascade in MHD turbulence proceeds from the Alfv\'enic stage to the ultimate resistive dissipation through a new, reconnection-mediated turbulent cascade, and the first theoretical studies of such a transition to the sub-inertial cascade were presented. 

To describe their results let us assume that MHD turbulence is driven at a large scale $L_0$ with the velocity $V_0$ which is also on the order of the large-scale Alfv\'en velocity~$V_{A0}$. According to the picture developed in \cite[][]{boldyrev2005,boldyrev_spectrum_2006}, the turbulent eddies can be viewed as current sheets with the dimensions $\lambda$ and $\xi$ in the plane normal to the local guide field, and $\ell$ along the field. These scales are related as\footnote{{These expressions assume, without loss of generality, that the fluctuations are isotropic at the outer scale of turbulence, that is, 
$\ell\sim\xi\sim\lambda$ at $\lambda\sim L$. Our results can be easily generalized to the case when the outer-scale velocity field is not comparable to $V_{A0}$, in which case the outer-scale Alfv\'enic Mach number will enter the results. This, however, will not change any relevant scaling relations in our theory.}}
\begin{eqnarray}
\xi\sim L_0\left({\lambda}/{L_0}\right)^{3/4},\label{xi}\\
\ell\sim L_0\left({\lambda}/{L_0}\right)^{1/2}.\label{ell}
\end{eqnarray}  
The magnetic and velocity fluctuations in such an eddy are then aligned in the field-perpendicular plane within a small, scale-dependent angle
\begin{eqnarray}
\theta\sim \lambda/\xi\sim \left({\lambda}/{L_0} \right)^{1/4}.
\label{theta}
\end{eqnarray}
As a result of a constant energy cascade toward small scales, the magnetic and velocity fluctuations in the inertial range scale as 
$v_{\lambda}\sim v_{A\lambda}\sim V_{A0}\left(\lambda/L_0 \right)^{1/4}$,
leading to the MHD energy spectrum
\begin{eqnarray}
E(k_\perp)d k_\perp \propto k_\perp^{-3/2}dk_\perp ,
\end{eqnarray}
\citep[e.g.,][]{maron_g01,haugen_04,muller_g05,mininni_p07,chen_11,mason_cb06,mason_cb08,mason2011,mason2012,perez_b10_2,perez_etal2012,perez_etal2014,chandran_intermittency_2015}.
As can be seen from~(\ref{theta}), at smaller scales the current sheets become progressively thinner, so their tearing instability becomes increasingly more important.  As demonstrated in \cite[][]{loureiro2017,mallet2017}, the fastest growing tearing mode in such a current sheet is the so-called Coppi mode with the growth rate $\gamma_t\sim (v_\lambda/\lambda)(v_\lambda \lambda/\eta)^{-1/2}$ \cite[][]{FKR,coppi_resistive_1976}. A formal comparison of the reconnection growth rate and the rate of nonlinear interaction (the eddy turnover rate) suggests that at the scale 
\begin{eqnarray}
\lambda_c\sim L_0 (L_0 V_{A0}/\eta)^{-4/7},
\label{lambda_c}
\end{eqnarray}
the tearing instability rate becomes comparable to the eddy turnover rate  
\begin{eqnarray}
\gamma_{nl}\sim v_{\lambda}\theta/\lambda
\label{gamma_nl}
\end{eqnarray} 
\cite[][]{loureiro2017,mallet2017}. At scales smaller than the critical scale ${\lambda_c}$ the nature of the MHD turbulence changes, as the interaction becomes mediated by the tearing instability and magnetic reconnection.\footnote{{This statement should be understood in a statistical sense. Not every current sheet formed by turbulence is necessarily reconnecting. We only claim that reconnection events become statistically significant enough to change the spectrum of turbulence at scales $\lambda < \lambda_c$.}} The goal of the present work is to describe the structure and scaling of MHD turbulence in this interval.

\section{Tearing instability: dimensional analysis} 
\label{sec:dimensional}
According to the theory of the tearing instability \cite[]{FKR,coppi_resistive_1976}, the spatial structure of the tearing mode has three characteristic scale parameters. The first scale parameter, $1/\Delta'$, characterizes the small scale structure developed by the outer solution, that is, the solution at scales not affected by the resistivity. The second scale parameter, $\delta_{in}$, is the resistive inner scale below which the mode structure is defined by the resistive diffusion. The fastest growing tearing mode, termed the Coppi mode in the analysis of \cite[][]{uzdensky_magnetic_2014}, corresponds to the case where $\delta_{in}\sim 1/\Delta'$, and, therefore, this mode is characterized by a single length scale, which we simply denote by~$\delta$ ($\delta\sim \delta_{in}\sim 1/\Delta'$). 

The Coppi  mode has the scale $\delta$ in the direction across the current layer (which we choose as the $x$ direction) and the scale $\zeta$ along it (the $y$ direction). The scale $\zeta$ is not independent in the Coppi mode, but is related to $\delta$. The analytical treatment is simplified in the regime $\zeta \gg \delta$, which we will  generally imply, but expect  that our final results may be extrapolated, at least dimensionally, to $\zeta\gtrsim \delta$. 
The third scale parameter is the width of the tearing mode island, $w$, which depends on the amplitude of the mode. For the linear tearing mode this width should be small, $w\ll 1/\Delta'$. We however will be interested in the early nonlinear regime, which corresponds to $w\sim 1/\Delta'$. This means that the {\em only} scale parameter characterizing the Coppi mode in the early nonlinear regime is the scale~$\delta$.
 
The spatial structure of the mode can be understood (at scales larger than $\delta$) from the force balance condition $-\nabla p+{\bf j}\times{\bf B}= 0$. Indeed, the evolution of the mode happens on the resistive time scale $\delta^2/\eta$, while the force balance is established on the Alfv\'enic time associated with the background profile. We denote this background field, which is directed along ${\hat y}$ and varies in the $x$ direction at scale $\lambda$,  by ${\bf B}_{\lambda}(x)$. We assume that this background field evolves slower than the tearing mode, keeping in mind that in the following sections we would like to extrapolate the final results to the most interesting case when the evolution times are comparable.

The tearing-mode field, which we denote ${\delta \bf b}_\lambda(x,y)$, changes in the $x$ and $y$ directions at the corresponding scales $\delta$ and $\zeta$. In Alfv\'enic units those fields are ${\bf v}_{A\lambda}$ and ${\delta \bf v}_{A\lambda}$. We then substitute ${\bf B}={\bf B}_{\lambda}+{\delta \bf b}_{\lambda}$ into the (curled) force-balance equation $\nabla \times \left[{\bf j}\times{\bf B}\right]=0$.

For the most unstable (Coppi) mode, linear theory gives:\footnote{In the literature on magnetic reconnection it is customary to denote the layer thickness $\lambda$ as $a$, and the mode dimension $\zeta$ as $1/k$.}
\begin{eqnarray}
\delta\sim \lambda^2/\zeta \label{delta}.
\label{relation}
\end{eqnarray}
In order to find the level of the field $\delta {\bf v}_{A\lambda}$ at which the mode becomes nonlinear, we balance the linear and nonlinear terms,
which leads to
\begin{eqnarray}
{\delta v}_{A\lambda,y}\sim {\delta v}_{A\lambda,x}\left(\zeta/\delta\right)\sim v_{A\lambda}\left(\delta/\lambda\right).\label{v_lambda}
\end{eqnarray}
Another view of Eq.~(\ref{v_lambda}) is the comparability of the tearing mode current ${\delta v}_{A\lambda,y}/\delta$ and the current of the background profile $v_{A\lambda}/\lambda$.

We now use the condition that the most unstable tearing mode evolves on the resistive time scale, which, as the tearing mode nears its nonlinear stage, also becomes comparable to its nonlinear evolution time:
\begin{eqnarray}
\gamma_t\sim \eta/\delta^2\sim {\delta v}_{A\lambda,x}/\delta .
\label{growth}
\end{eqnarray}
Remarkably, equations~(\ref{delta}, \ref{v_lambda}, \ref{growth}) allow us to express all the parameters of the nonlinear Coppi mode through the eddy scale~$\lambda$:  
\begin{eqnarray}
&\label{zeta_t} \zeta \sim \lambda \left(\lambda v_{A\lambda}/\eta \right)^{1/4},\\
&\label{delta_t}\delta \sim \lambda \left(\lambda v_{A\lambda}/\eta\right)^{-1/4}, \\
&\label{gamma_t} \gamma_t\sim (v_{A\lambda}/\lambda)(\lambda v_{A\lambda}/\eta)^{-1/2}, \\ 
& {\delta v}_{A\lambda,x}\sim v_{A\lambda}\left(\lambda v_{A\lambda}/\eta\right)^{-3/4},\\
& {\delta v}_{A\lambda,y}\sim v_{A\lambda}\left(\lambda v_{A\lambda}/\eta\right)^{-1/4}.
\end{eqnarray}
Our dimensional derivation of the most unstable tearing mode is important for the phenomenological analysis of the reconnection-mediated turbulence, which we will present in the following sections.

\section{Reconnection-mediated turbulence: A simple phenomenological model}
\label{simple}
In order to construct a model of turbulence constrained by the tearing instability of the eddies, we shall make two critical assumptions.

First, we observe that analytical and numerical results in the case of a slowly evolving, laminar reconnection profile show that as the tearing (Coppi) mode enters its nonlinear stage, the {\em whole} magnetic profile is distorted by it on the Alfv\'enic time scale \cite[e.g.,][]{waelbroeck_onset_1993,loureiro_X-point_2005}. Our assumption is that this remains true in the case at hand here, where the tearing mode is evolving on a turbulent, dynamic background.
  As the tearing mode enters the nonlinear stage, the turbulent eddy responds to it by adjusting its own eddy-turnover time to the evolution time of the mode.

The second assumption is that the tearing mode itself does not get past its early nonlinear stage, and the usual X-point collapse that pertains to the laminar situation \cite[]{waelbroeck_onset_1993,loureiro_X-point_2005} does not take place. This is because the timescale for the X-point collapse is Alfv\'enic, i.e., the same as that on which the turbulent background is evolving. The absence of a timescale separation therefore should preclude X-point collapse from occurring.\footnote{{This statement does not imply that X-point collapse and subsequent plasmoid formation do not happen in MHD turbulence. Such structures may occasionally be generated by a turbulent flow \cite[e.g.,][]{wan2013}. We only propose that such structures are not statistically significant for the energy spectrum in the considered interval.}}

In the phenomenology of \cite[][]{boldyrev_spectrum_2006}, the rate of nonlinear interaction within an anisotropic eddy is controlled by the alignment angle associated with the eddy $\gamma_{nl}\sim {v}_{A\lambda}\theta/\lambda$. In the Alfv\'enic cascade this angle is given by Eq.~(\ref{theta}). We propose that the nonlinear Coppi mode affects the evolution of the eddy by distorting the alignment angle of the magnetic lines. The typical distortion of the alignment angle in such a tearing mode is
\begin{eqnarray}
\theta_t\sim \delta/\zeta.
\label{theta_t}
\end{eqnarray}  
One can check from (\ref{zeta_t}), (\ref{delta_t}), and (\ref{gamma_t}), that in order for the nonlinear evolution time of the eddy to be comparable to that of the tearing mode, one needs to require that $\theta \sim \theta_t$. In other words the angular distortion provided by the nonlinear tearing mode affects the whole eddy of size $\lambda$.   In the Alfv\'enic regime $\lambda \gg\lambda_c$, the tearing distortion of the alignment angle is not essential: $\theta_t\ll \theta$. However, below the scale $\lambda_c$, the tearing distortion (\ref{theta_t}) dominates.

\section{Spectrum of reconnection-mediated turbulence}
\label{sec:spectrum}
According to our discussion in the previous section, we assume that at all the reconnection dominated scales $\lambda<\lambda_c$ the energy cascade is governed by the balance between the nonlinear eddy turnover time and the linear tearing time. The energy flux over scales then can be estimated as
\begin{eqnarray}
\label{cascade}
\gamma_{nl}v_{A\lambda}^2=\epsilon,   
\end{eqnarray}	
where $\epsilon$ is the constant rate of the energy cascade over scales. This estimate implies that reconnection is not leading to energy dissipation at the considered scales. This is in agreement with our  picture where the eddy is essentially destroyed by the nonlinear tearing mode on its dynamical time. In laminar tearing mode studies, the energy dissipation is only significant in the late nonlinear regime, after X-point collapse has taken place; here this does not happen, so there is little energy dissipation.

\begin{figure}
\includegraphics[width=\columnwidth,trim=0 0 0 0,clip]{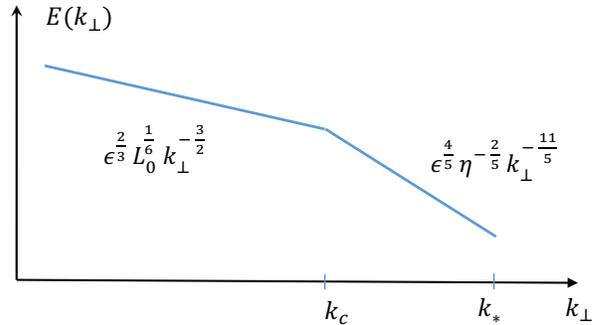}
\caption{Sketch of the energy spectrum (in a log-log scale). In the interval $k<k_c\sim 1/\lambda_c$, the spectrum is dominated by Alfv\'enic turbulence, while in the interval $k>k_c$ it is mediated by tearing instability.}
\label{fig:spectrum}
\end{figure}

One can estimate from the large-scale conditions, $\epsilon\sim V^3_{A0}/L_0$. From Eqs.~(\ref{gamma_t}) and (\ref{cascade}) we then obtain:
\begin{eqnarray}
v_{A\lambda}\sim \epsilon^{2/5}\eta^{-1/5}\lambda^{3/5}.
\end{eqnarray}    
This leads to the Fourier energy spectrum 
\begin{eqnarray}
E(k_\perp)dk_\perp \sim \epsilon^{4/5}\eta^{-2/5}k_\perp^{-11/5}d k_\perp.
\label{spectrum}
\end{eqnarray}
 Our model also allows us to derive the dissipation cutoff $k_*$ of the spectrum. Noting that the energy dissipation per unit time is given by
\begin{eqnarray}
\eta \int\limits^{k_*}k_\perp^2 E(k_\perp)\,dk_\perp \sim\epsilon, 
\end{eqnarray}
we obtain\footnote{A similar scaling of the energy and the dissipation cutoff has been recently proposed by \cite[][]{mallet2017} based on a dynamical picture that is qualitatively different and complementary to ours, see section~\ref{Discussion}. } 
\begin{eqnarray}
k_*\sim \epsilon^{1/4}\eta^{-3/4} \sim L_0^{-1} S_0^{3/4}.
\label{k*}
\end{eqnarray}	
It can be checked that at the dissipation scale $\lambda_*\sim 1/k_*$ the local Lundquist number is $S_{\lambda_*}\equiv \lambda_* v_{A\lambda_*}/\eta\sim 1$. The behavior of the energy spectrum in both the Alfv\'enic and reconnection-mediated regimes is illustrated in Fig.~(\ref{fig:spectrum}). 

The spectrum (\ref{spectrum}) is slightly shallower than the spectrum $k_{\perp}^{-5/2}$ proposed for the reconnection-mediated interval in~\cite[][]{loureiro2017}. It is instructive to see where the difference comes from. The estimate of \cite[][]{loureiro2017} follows from the picture that as the magnetic profile at the critical scale $\lambda_c$ becomes unstable, it triggers an X-point collapse during which the reconnecting field $v_{A\lambda_c}$ does not significantly change~\cite[][]{loureiro_X-point_2005}. This would be true for the tearing instability initiated on a slowly evolving background \cite[e.g.,][]{uzdensky_fast_2010,loureiro_magnetic_2012,uzdensky_magnetic_2014}.
Upon  approximating  the velocity $v_{A\lambda}$ in the instability rate~(\ref{gamma_t}) by a scale-independent velocity $v_{A\lambda_c}$, one formally re-derives the scaling~$-5/2$.  

The self-consistent model discussed in this section takes into account the fact that the reconnection is initiated not on a slowly evolving background, but rather on a dynamic background evolving on the same Alfv\'enic time scale. As a result, the X-point collapse does not have a chance to set in. Rather, the tearing instability leads to creation of even smaller eddies thus mediating the turbulent cascade.  This dynamic picture requires one to use the {\em scale-dependent} velocity $v_{A\lambda}$ in the eddy turnover rate~(\ref{gamma_t}), which leads to the spectrum~(\ref{spectrum}).

\section{Anisotropy of the reconnection-mediated turbulence}
\label{sec:anisotropy}
\begin{figure}
\includegraphics[width=\columnwidth,trim=0 0 0 0,clip]{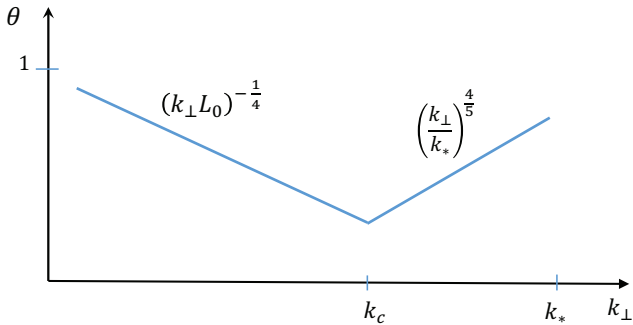}
\caption{Sketch of the alignment angle in a log-log scale as a function of $k\sim 1/\lambda$. For $k<k_c$, the angle decreases and the turbulent eddies become progressively more anisotropic as $k$~approaches $k_c$. For $k>k_c$, the alignment angle increases, the eddies become progressively more isotropic as $k$~approaches the dissipation scale~$k_*$.}
\label{fig:angle}
\end{figure}

In order to study the anisotropy, it is instructive to analyze the behavior of the alignment angle $\theta$ in both the Alfv\'enic and reconnection-mediated regions. In the Alfv\'enic interval, $\lambda > \lambda_c$, the alignment angle is given by Eq.~(\ref{theta}). It decreases with decreasing scale until the reconnection scale $\lambda_c$ is reached. In the reconnection-mediated interval, $\lambda<\lambda_c$, the behavior of the alignment angle changes. According to Eq.~(\ref{theta_t}), the angle is now increasing with decreasing scale. We summarize this behavior as follows:
\begin{eqnarray}
&\theta\sim \left({\lambda}/{L_0} \right)^{1/4}, \quad \lambda>\lambda_c ; \\
&\theta\sim \left({\lambda}/{\lambda_*} \right)^{-4/5}, \quad \lambda_*< \lambda<\lambda_c. \label{theta_tearing}
\end{eqnarray} 
This behavior is schematically illustrated in Fig.~(\ref{fig:angle}). At the dissipation scale $\lambda_*\sim 1/k_*$ the alignment becomes lost, meaning that the eddy sizes in the guide-field-perpendicular direction, $\lambda$ and $\xi$, become comparable to each other. 
To find the eddy size in the field-parallel direction, we note, following the standard argument \cite[][]{goldreich_toward_1995,boldyrev2005,boldyrev_spectrum_2006}, that during the nonlinear evolution time the turbulent fluctuations get correlated along the background magnetic field at the scale $\ell\sim V_{A0}/\gamma_{nl}$. This allows us to find the sizes of eddies as
\begin{eqnarray}
&\xi\sim \lambda/\theta \sim L_0(\lambda_c/L_0)^{1/4}(\lambda/\lambda_c)^{9/5}, \label{xi_align}\\ 
&\ell\sim L_0 (\lambda_c/L_0)^{1/2} (\lambda/\lambda_c)^{6/5},\label{ell_align}
\end{eqnarray}
which extends the results (\ref{xi}) and (\ref{ell}) into the reconnection-mediated region. The turbulent eddies assume the dimensions $\xi_* \sim \lambda_* \ll \ell_*$ as their scale approaches the dissipation scale, where $\ell_*\sim L_0S_0^{-1/5}$. Their shapes approach that of filaments, or current ropes, oriented along the direction of the local large-scale magnetic field.

Our results also explain why the scaling of the dissipation cutoff $k_*$ coincides with the scaling of the Kolmogorov  cutoff in nonmagnetized turbulence. Indeed, when the alignment and the corresponding reduction of the nonlinear interaction are lost, the estimates of the nonlinear interactions time and of the dissipation scale dimensionally coincide with those of nonmagnetized turbulence, cf.~\cite{goldreich_toward_1995,boldyrev_spectrum_2006,perez_etal2014a,perez_etal2014}.

\section{Non-universality of reconnection-mediated turbulence} In contrast with Alfv\'enic turbulence existing at $\lambda >\lambda_c$, the reconnection-mediated turbulence is non-universal, in that it depends on the mechanism of resistive dissipation. To  illustrate this, we consider the hyper-resistivity of order~$s$ provided in the Fourier space by the operator ${\tilde \eta}k^{2+2s}$; the regular resistivity is recovered at $s=0$.\footnote{The hyper-resistive tearing mode of order $s=1$ has been studied before \cite[e.g.,][]{aydemir1990,huang2013}. The results of our dimensional derivation for $s=1$ agree with the analytical solution obtained in these works.} In the hyper-resistive case, our basic equations (\ref{delta}) and (\ref{v_lambda}) remain intact. Eq.~(\ref{growth}), however, is replaced by  
\begin{eqnarray}
\gamma_t\sim {\tilde\eta}/\delta^{2+2s}\sim {\delta v}_{A\lambda,x}/\delta .
\label{growth_hyper}
\end{eqnarray}
From Eqs.~(\ref{delta}), (\ref{v_lambda}), and (\ref{growth_hyper}) we then derive the parameters of the fastest growing mode as:
\begin{eqnarray}
&\zeta \sim \lambda {\tilde S}_\lambda^{{1}/{(2(2+s))}},\\
&\delta \sim \lambda {\tilde S}_\lambda^{-{1}/{(2(2+s))}}, \\
&\label{gamma_t_hyper} \gamma_t\sim \left({v_{A\lambda}}/{\lambda}\right){\tilde S}_\lambda^{-{1}/{(2+s)}}, \\ 
& {\delta v}_{A\lambda,x}\sim v_{A\lambda} {\tilde S}_\lambda^{-{3}/{(2(2+s))}},\\
& {\delta v}_{A\lambda,y}\sim v_{A\lambda} {\tilde S}_\lambda^{-{1}/{(2(2+s))}},
\end{eqnarray}
where ${\tilde S}_\lambda=v_{A\lambda}\lambda^{1+2s}/{\tilde \eta}$ is the hyper-resistive Lundquist number at scale~$\lambda$. The transition to the reconnection-mediated regime occurs at the scale where the rate (\ref{gamma_t_hyper}) becomes comparable to the Alfv\'enic eddy turnover rate~(\ref{gamma_nl}), which gives the transition scale
\begin{eqnarray}
\lambda_c\sim L_0 {\tilde S}_0^{-{4}/{(7+9s)}},
\label{lambda_c_hyper}
\end{eqnarray}
where the outer-scale Lundquist number is defined as ${\tilde S}_0=v_{A0}L_0^{1+2s}/{\tilde \eta}$. The derivation of the energy spectrum and the corresponding eddy anisotropy is then completely analogous to our discussion in the preceding sections, which gives:
\begin{eqnarray}
E(k_\perp)\sim \epsilon^{{(4+2s)}/{(5+3s)}}{\tilde\eta}^{-{2}/{(5+3s)}}k_\perp^{-{(11+9s)}/{(5+3s)}},
\end{eqnarray}
and
\begin{eqnarray}
\xi\sim \lambda/\theta, \quad  \mbox{where} \quad \theta\sim(\lambda/\lambda_*)^{-{(4+6s)}/{(5+3s)}}.
\end{eqnarray}
The dissipation scale coincides with that in the hyper-viscous Kolmogorov phenomenology
\begin{eqnarray}
\lambda_* \sim 1/k_* \sim L_0{\tilde S}_0^{-{3}/{(4+6s)}}\sim \epsilon^{-{1}/{(4+6s)}}{\tilde\eta}^{{3}/{(4+6s)}},
\label{k*_hyper}
\end{eqnarray}  
and the eddies turn into filaments at the dissipation scale.

In order for the reconnection-mediated interval to be observed in numerical simulations, the tearing scale (\ref{lambda_c}) and the dissipation scale (\ref{k*}) should be well separated, say by an order of magnitude. For that one needs the Lundquist number $S_0>10^{5.6}$, which is a challenge for the present-day numerical simulations. This restriction becomes even more prohibitive for the systems with hyper-resistivity. From (\ref{lambda_c_hyper}) and (\ref{k*_hyper}) we derive 
\begin{eqnarray}
\left(\lambda_c/\lambda_* \right)\sim \left(L_0/\lambda_* \right)^{(5+3s)/(21+27s)}.
\end{eqnarray}
This means that for a given ratio of the outer scale $L_0$ and the dissipation scale $\lambda_*$, the ratio of the $\lambda_c$ and $\lambda_*$ decreases as the order of hyper-resistivity increases, making observation of the reconnection-mediated interval in hyper-resistive numerical simulation more difficult.

\section{Reconnection-mediated turbulence: a refined model}
\label{refined}
The simple model discussed in the previous sections has two important ingredients that, we believe, should survive in more refined treatments of reconnection-mediated turbulence. First is the assumption that the dynamics at the reconnection-dominated scales should depend on a {\em single} scale parameter -- the dissipation scale $\lambda_*$, see e.g..~(\ref{theta_tearing}). This assumption then requires that the dissipation scale has the Kolmogorov-like form~(\ref{k*}). The second assumption is that in the reconnection-dominated regime, the tearing time and the nonlinear Alfv\'enic time are on the same order, so that the X-point collapse and saturation do not occur.   

The assumption that requires a revision is the assumption that the tearing mode grows at the time scale dictated by the molecular magnetic diffusivity~$\eta$. Indeed, as one can check, the inner scale developed by the tearing mode always exceeds the dissipation scale of turbulence, that is, $\delta \geq \lambda_*$. This implies that in order to treat the inner structure of the mode properly one needs to use a ``renormalized'', turbulent diffusivity that is larger than the molecular diffusivity. To implement this in our model we note that larger diffusivity leads to a larger tearing growth rate. We may therefore assume that the tearing mode operating on a dynamic turbulent background leads to the growth rate
\begin{eqnarray}
\gamma_t\sim \left(v_{A\lambda}/\lambda \right)\theta_t,
\end{eqnarray}    
with the alignment angle 
\begin{eqnarray}
\theta_t\sim \left(\lambda/\lambda_* \right)^{-4/5+\beta},
\label{theta_refined}
\end{eqnarray}
where $0\leq  \beta < 4/5$. The growth rate of the ``classical" tearing mode operating on a slow laminar background would then formally correspond to $\beta=0$, as we discussed in the previous sections. Currently, a more detailed theory of the reconnection-dominated turbulence is not available, so the evaluation of the parameter $\beta$ should await further analytical and numerical studies. 

The corresponding Fourier energy spectrum is then 
\begin{eqnarray}
\label{E_refined}
E(k_\perp)\propto k_\perp^{-11/5+2\beta/3}, 
\end{eqnarray}
while the transition scale is 
\begin{eqnarray}
\label{lambda_refined}
\lambda_c/L \propto S_L^{-(4-5\beta)/(7-{20\beta/3})}. 
\end{eqnarray}
The refined model of reconnection-mediated turbulence therefore predicts a shallower energy spectrum and, importantly, a larger scale of transition as compared to our simple model described in the previous sections. 

{Finally, one can envision a modification of our simple model developed in sections~\ref{sec:dimensional} -- \ref{sec:anisotropy}, which may stem from assuming different magnetic profiles of the eddies. In our treatment in section~\ref{sec:dimensional} the magnetic profile $v_{A\lambda}(x)$ was assumed to qualitatively resemble the $\tanh(x/\lambda)$ profile often discussed in the reconnection literature. One can however envision less trivial profiles of the reconnecting magnetic field, for instance, resembling that of $\sin(x/\lambda)$. The only modification required in this case is the replacement of Eq.~(\ref{relation}) by $\delta\sim \lambda^3/\zeta^2$. This however would change the results only slightly. The transition scale~(\ref{lambda_c}) would be changed to $\lambda_c\sim L_0S_0^{-6/11}$, while the energy spectrum~(\ref{spectrum}) to $E(k_\perp)\sim k_\perp^{-19/9}$.\footnote{On a purely numerical basis, one notices that this solution is also recovered from the general formulae (\ref{E_refined}) and (\ref{lambda_refined}), at~$\beta=2/15$.} }

\section{Discussion and Conclusion}
\label{Discussion}
We have proposed a model for reconnection-mediated MHD turbulence, a regime discovered in recent works by \citet{loureiro2017,mallet2017}. Our derivation is based on the scale-dependent dynamic alignment of turbulent fluctuations in the guide-field-perpendicular direction, given by Eqs.~(\ref{theta_tearing}) and (\ref{theta_refined}). It extends the theory of scale-dependent dynamic alignment in Alfv\'enic turbulence into the reconnection-mediated interval. 

The scalings~(\ref{spectrum}) and (\ref{ell_align}) coincide with one of the solutions proposed in \cite[][]{mallet2017} based on modeling of a turbulent field as a discrete hierarchy of current sheets undergoing a succession of X-point collapses and on applying a coarse-graining procedure to obtain the spectrum, an approach different from ours. {The scale-dependent dynamic alignment (\ref{theta_tearing}) and (\ref{theta_refined}) is not derived in their model.} The coincidence of the spectra is not surprising, however, since both models predict the same dissipation cutoff~(\ref{k*}). Indeed, the energy scaling (\ref{spectrum}) can be easily obtained if one accepts that the turbulent structures at the dissipation cutoff are filamentary current ropes stretched along the local guide field. Such structures imply the absence of the dynamic alignment, and as a result, the Kolmogorov-like scaling of the dissipation cutoff~(\ref{k*}). Following \cite[][]{loureiro2017}, 
one then writes the general power-law spectrum in the reconnection-mediated interval as 
\begin{eqnarray}
E(k)dk\propto (k/k_c)^{-\alpha}k_c^{-3/2}dk,
\end{eqnarray} 
{where $k_c\propto \eta^{-4/7}$}. The requirement that the rate of energy  dissipation in the turbulent cascade, $\epsilon=\int^{k_*} E(k) \eta k^2 dk$, is independent of $\eta$, then gives~$\alpha=11/5$.

We however caution that a mere observation of the Kolmogorov-like scaling of the small-scale cutoff {(without an observation of the reconnection-mediated inertial interval)} in numerical simulations does not automatically imply the presence of the reconnection-mediated cascade. Indeed, the scale-dependent alignment is always lost deep in the dissipation region, no matter what the Lundquist number is \cite[e.g.,][]{perez_etal2012}. In addition, as demonstrated in \cite[][]{perez_etal2014a,perez_etal2014}, the alignment can be easily broken in simulations by purely numerical effects, such as proximity to the dealiasing cutoff or lack of numerical resolution at small scales. These effects similarly lead to the Kolmogorov-like scaling of the spectral cutoff.

We may however compare our results with the numerical simulations of turbulence generated inside a reconnection layer by \cite[][]{huang_turbulent_2016}. In their set up the reconnection layer is not formed by turbulence, but rather imposed as a large-scale condition. {The X-point collapse and plasmoid formation are observed at the initial stages of the evolution. Those structures, however, do not appear to be pronounced in the fully developed turbulent regime, which seems to be consistent with our picture.}
One may, therefore, expect that some features of reconnection-mediated turbulence may be present in the simulations of \citet{huang_turbulent_2016}. 
They found that such turbulence has the spectrum of magnetic fluctuations $E(k_\perp)\propto k_\perp^{-2.1}$ \dots $k_\perp^{-2.3}$, which agrees with our predictions.\footnote{Their spectrum of the kinetic fluctuations was a little steeper though, with the spectral exponent in the range from $-2.3$ to $-2.5$. We however note that the simulations of \cite[][]{huang_turbulent_2016} are compressible, in which case there is no universal way of defining the velocity variable that should exhibit a universal scaling behavior \cite[e.g.,][]{kritsuk2007}}


\paragraph{Acknowledgments.}
{We thank the referee, Alex Schekochihin, for his careful reading of our manuscript and for useful remarks.} SB is partly supported by the National Science Foundation under the grant NSF AGS-1261659 and by the Vilas Associates Award from the University of Wisconsin - Madison. NFL was supported by the NSF-DOE Partnership in Basic Plasma Science and Engineering, award no. DE-SC0016215. 



\end{document}